\documentclass[aps,amsfonts,amsmath,prd,preprint,nofootinbib]{revtex4}

\begin{document}

\title{The vacuum energy crisis}

\author{Alexander Vilenkin}

\address{Institute of Cosmology, Department of Physics and Astronomy,\\
Tufts University, Medford, MA 02155, USA}

\begin{abstract}

The smallness of the vacuum energy density and its near coincidence
with the average matter density of the universe are naturally
explained by anthropic selection. An alternative explanation, based on
the cyclic model of Steinhardt and Turok, does not address the
coincidence problem and is therefore less convincing. This article
appeared in ``Science'' (4 May 2006) as a ``perspective'' for
Steinhardt and Turok's paper in the same issue (astro-ph/0605173).

\end{abstract}

\maketitle

One of the stranger consequences of quantum mechanics is that even
empty space has energy.  The problem of how to calculate this vacuum
energy is arguably the most intriguing mystery in theoretical
physics. For decades physicists tried to understand why this energy is
so small, but no definitive solution has yet been found. In this
week's {\it Science} Express, Steinhardt and Turok propose a new
approach \cite{ST}.

Vacuum is empty space, but it is far from being ``nothing''. It is a
complicated physical object in which particles like electrons,
positrons and photons are being incessantly produced and destroyed by
quantum fluctuations. Such virtual particles exist only for a fleeting
moment, but their energies combine to endow each cubic centimeter of
space with a nonzero energy. This vacuum energy density does not
change in time; it is called the cosmological constant and is usually
denoted by $\Lambda$.  The trouble is that theoretical calculations of
$\Lambda$ give ridiculously large numbers, 120 orders of magnitude
greater than what is observed. According to Einstein's General
Relativity, vacuum energy produces a repulsive gravitational force,
and if the energy were so large, its gravity would have instantly
blown the universe apart.

It is conceivable that positive vacuum energy contributions from
some particle species are compensated by negative contributions
from other species, so that the net result is close to zero.  But then
the compensation must be amazingly precise, up to 120 decimal
places. There seems to be no good reason for such a miraculous
cancellation. Until recently, the great majority of physicists
believed that something so small could only be zero: some hidden
symmetry should force exact cancellation of all contributions to the
cosmological constant. However, observations of distant supernova
explosions in the late 1990's yielded the surprising discovery that
the expansion of the universe accelerates with time \cite{supernova}
-- a telltale sign of cosmic repulsion caused by a nonzero
(positive) cosmological constant.

The observed magnitude of $\Lambda$ has brought about another mystery:
its value is roughly twice the average energy density (or, equivalently,
mass density) of matter in the universe. This is surprising because
the matter and vacuum densities behave very differently with cosmic
expansion. The vacuum density remains constant, while the matter
density decreases; it was much greater in the past and will be much
smaller in the future.  Why, then, do we happen to live at the very
special epoch when the two densities are so close to one another? This
became known as the cosmic coincidence problem.

Both puzzles can be resolved if one is prepared to assume that the
cosmological constant is not a fixed number, but takes a wide variety
of values in remote parts of the universe. In regions where it is much
larger than the observed value, its repulsive gravity will be stronger
and will prevent matter from clumping into galaxies and stars
\cite{Weinberg,Linde}. Life is not likely to evolve in such regions.

The idea of ``anthropic selection'' -- that certain features of the
universe are selected by the requirement that observers should be
there to detect them \cite{LR} -- runs contrary to the physicist's
aspiration to derive all constants of nature from first principles. It
has been passionately resisted by the physics community, but has
recently gained support from both string theory and cosmology.  String
theory, the most promising candidate for the fundamental theory of
nature, predicts a multitude of vacuum states characterized by
different values of $\Lambda$ and other ``constants''. Inflationary
cosmology, which now has a substantial observational support, suggests
that the universe on the largest scales is in a state of high-energy
exponential expansion and is constantly spawning low-energy
``bubbles'' like ours, with all possible values of the ``constants''
(see Fig.1). Galaxies and observers exist only in rare bubbles where
$\Lambda$ is small and other constants are also appropriately
selected. Analysis shows that most of the galaxies are formed in
regions where vacuum and matter densities are about the same at the
epoch of galaxy formation \cite{AV,Efstathiou,MSW}. Our present time
is close to that epoch, and this explains the coincidence
\cite{GLV,Bludman}.

Steinhardt and Turok propose an alternative explanation for the
smallness of $\Lambda$. Building on the idea of Abbott \cite{Abbott},
they postulate the existence of a long sequence of vacuum states with
$\Lambda$ changing in small increments from one state to the next. If
the universe starts with a large cosmological constant, its value will
be gradually reduced through a sequence of quantum transitions to
lower and lower values. Abbott showed that as $\Lambda$ approaches
zero, the transitions become increasingly slow, so the universe spends
most of the time in the state with the smallest positive $\Lambda$. He
found, however, that the descent to small values of $\Lambda$ takes so
long that all matter gets completely diluted by the cosmic expansion,
and one ends up with an empty universe. To fix this flaw, Steinhardt
and Turok combine Abbott's model with the cyclic cosmological scenario
\cite{cyclic}, in which the universe goes through multiple cycles of
expansion and contraction. The high density of matter is regenerated
at the start of each cycle, so `the empty universe problem' is
solved. Most of the cycles will occur while the universe is in the
lowest-$\Lambda$ state, and Steinhardt and Turok argue that this state
is most likely to be observed.

The cyclic model is still being developed and is not widely
accepted. More importantly, although the S\&T proposal may explain the
smallness of $\Lambda$, it does not address the cosmic coincidence
problem: why should the smallest possible value of $\Lambda$ be
comparable to the present matter density of the universe?  The
anthropic explanation appears, therefore, to be more compelling.

~~~

This work was supported in part by the National Science Foundation.

\end{document}